\newcommand{\be}{\begin{equation}}\newcommand{\ee}{\end{equation}}
\newcommand{\bea}{\begin{eqnarray}}\newcommand{\eea}{\end{eqnarray}}
\documentstyle[12pt]{article}
\def\PRL #1 #2 #3{{\sl Phys. Rev. Lett.} {\bf#1} (#2) #3}
\def\NPB #1 #2 #3{{\sl Nucl. Phys.} {\bf B#1} (#2) #3}
\def\NPBFS #1 #2 #3 #4{{\sl Nucl. Phys.} {\bf B#2} [FS#1] (#3) #4}
\def\CMP #1 #2 #3{{\sl Commun. Math. Phys.} {\bf #1} (#2) #3}
\def\PRD #1 #2 #3{{\sl Phys. Rev.} {\bf D#1} (#2) #3}
\def\PLA #1 #2 #3{{\sl Phys. Lett.} {\bf #1A} (#2) #3}
\def\PLB #1 #2 #3{{\sl Phys. Lett.} {\bf #1B} (#2) #3}
\def\JMP #1 #2 #3{{\sl J. Math. Phys.} {\bf #1} (#2) #3}
\def\PTP #1 #2 #3{{\sl Prog. Theor. Phys.} {\bf #1} (#2) #3}
\def\SPTP #1 #2 #3{{\sl Suppl. Prog. Theor. Phys.} {\bf #1} (#2) #3}
\def\AoP #1 #2 #3{{\sl Ann. of Phys.} {\bf #1} (#2) #3}
\def\PNAS #1 #2 #3{{\sl Proc. Natl. Acad. Sci. USA} {\bf #1} (#2) #3}
\def\RMP #1 #2 #3{{\sl Rev. Mod. Phys.} {\bf #1} (#2) #3}
\def\PR #1 #2 #3{{\sl Phys. Reports} {\bf #1} (#2) #3}
\def\AoM #1 #2 #3{{\sl Ann. of Math.} {\bf #1} (#2) #3}
\def\UMN #1 #2 #3{{\sl Usp. Mat. Nauk} {\bf #1} (#2) #3}
\def\FAP #1 #2 #3{{\sl Funkt. Anal. Prilozheniya} {\bf #1} (#2) #3}
\def\FAaIA #1 #2 #3{{\sl Functional Analysis and Its Application} {\bf
#1} (#2) #3}
\def\BAMS #1 #2 #3{{\sl Bull. Am. Math. Soc.} {\bf #1} (#2)
#3} \def\TAMS #1 #2 #3{{\sl Trans. Am. Math. Soc.} {\bf #1} (#2) #3}
\def\InvM #1 #2 #3{{\sl Invent. Math.} {\bf #1} (#2) #3}
\def\LMP #1 #2 #3{{\sl Letters in Math. Phys.} {\bf #1} (#2) #3}
\def\IJMPA #1 #2 #3{{\sl Int. J. Mod. Phys.} {\bf A#1} (#2) #3}
\def\AdM #1 #2 #3{{\sl Advances in Math.} {\bf #1} (#2) #3}
\def\RMaP #1 #2 #3{{\sl Reports on Math. Phys.} {\bf #1} (#2) #3}
\def\IJM #1 #2 #3{{\sl Ill. J. Math.} {\bf #1} (#2) #3}
\def\APP #1 #2 #3{{\sl Acta Phys. Polon.} {\bf #1} (#2) #3}
\def\TMP #1 #2 #3{{\sl Theor. Mat. Phys.} {\bf #1} (#2) #3}
\def\JPA #1 #2 #3{{\sl J. Physics} {\bf A#1} (#2) #3}
\def\JSM #1 #2 #3{{\sl J. Soviet Math.} {\bf #1} (#2) #3}
\def\MPLA #1 #2 #3{{\sl Mod. Phys. Lett.} {\bf A#1} (#2) #3}
\def\JETP #1 #2 #3{{\sl Sov. Phys. JETP} {\bf #1} (#2) #3}
\def\JETPL #1 #2 #3{{\sl  Sov. Phys. JETP Lett.} {\bf #1} (#2) #3}
\def\PHSA #1 #2 #3{{\sl Physica} {\bf A#1} (#2) #3}
\def\CQG #1 #2 #3{{\sl Class. Quantum Grav.} {\bf #1} (#2) #3}
\def\SJNP #1 #2 #3{{\sl Sov. J. Nucl. Phys. (Yadern.Fiz.)} {\bf #1} (#2) #3}
\def\a{\alpha}\def\b{\beta}\def\g{\gamma}\def\d{\delta}\def\e{\epsilon}

\def\k{\kappa}
\def\Th{\Theta}\def\th{\theta}\def\Om{\Omega}\def\G{\Gamma}
\newcommand{\nn}{\nonumber\\}\newcommand{\p}[1]{(\ref{#1})}
\begin{document}
\renewcommand{\thefootnote}{\fnsymbol{footnote}}
\thispagestyle{empty}
\begin{flushright}
Preprint DFPD 95/TH/06\\
February 1995
\end{flushright}

\bigskip
\bigskip
\begin{center}
{\large\bf
ON THE GENERALIZED ACTION PRINCIPLE FOR SUPERSTRINGS AND SUPERMEMBRANES
}\footnote{Work supported  in  part  by  the
International Science Foundation under the grant N RY 9000,
 by the State  Committee  for  Science  and  Technology  of
Ukraine under the Grant N 2/100 and
 by the INTAS grants 93--127, 93--493, 93--633}

\vspace{1cm}
{\bf Igor A. Bandos,}

\vspace{0.2cm}
{\it Kharkov Institute of Physics and Technology}
{\it 310108, Kharkov,  Ukraine}\\
e-mail:  kfti@kfti.kharkov.ua

\vspace{0.5cm}
\renewcommand{\thefootnote}{\dagger}
{\bf Dmitrij Sorokin\footnote{on leave from Kharkov Institute of
Physics and Technology, Kharkov, 310108, Ukraine.\\e--mail:
sorokin@pd.infn.it}},

\vspace{0.2cm}
{\it Universit\`a Degli Studi Di Padova
Dipartimento Di Fisica ``Galileo Galilei''\\
ed INFN, Sezione Di Padova
Via F. Marzolo, 8, 35131 Padova, Italia}

\vspace{0.3cm}
{\bf and}

\vspace{0.3cm}

{\bf Dmitrij V. Volkov}

\vspace{0.2cm}
{\it Kharkov Institute of Physics and Technology}
{\it 310108, Kharkov,  Ukraine}\\
e-mail:  dvolkov@kfti.kharkov.ua

\vspace{0.5cm}
%\vspace{0.3cm}
{\bf Abstract}
\end{center}

\medskip
We revise the twistor--like superfield approach to describing
super--p--branes by use of the basic principles of the
group--manifold approach \cite{rheo}. A super--p--brane action is
constructed solely of geometrical objects
as the integral over a (p+1)--surface. The Lagrangian is the external product
of supervielbein differential forms in world supersurface and target
superspace without any use of  Lagrange multipliers. This allows one to
escape the problem of infinite irreducible symmetries and redundant
propagating fields.  All the constraints on the geometry of
world supersurface and the conditions of its embedding into target
superspace arise from the action as differential form equations.

\setcounter{page}1
\renewcommand{\thefootnote}{\arabic{footnote}}
\setcounter{footnote}0

\newpage
\section{Introduction}
There are two main approaches to describing supersymmetric theories. One
is based on the $x$--space component formulation and another is the
superspace formalism, each of the approaches having its own advantages
and  drawbacks. The $x$--space component formulation explores the
minimal number of fields, but, as a rule, their local supersymmetry
transformation law is not easy to determine, and the off shell
formulation requires the introduction of auxiliary fields. The
superspace approach has an advantage of being manifestly
superdiffeomorphism invariant. The essential drawback of the
superspace approach is the necessity of introducing superfield
constraints whose geometrical meaning is sometimes very obscure and
which put some extended supersymmetry theories on the mass shell.

There are also a number of papers \cite{rheo} devoted to the
development of a new, so called, group--manifold approach to
supersymmetric field theory aimed to accumulate the
advantages of both conventional formulations and to get rid of their
drawbacks. The backbone of the group--manifold approach is a generalized
action and variation principle associated with another concept
(rheonomy) which substitutes the notion of supersymmetry invariance.
%Rheonomy is a mathematical concept which generalizes the idea of the
%horizontality of the curvatures on fiber bundles and allows one to
%provide the $x$--space with a special status in superspace.

In the present paper we propose an analogous approach to the theory of
superstrings and supermembranes. The reader well acquainted with the
papers \cite{rheo} can easily see that every notion and conjecture of
\cite{rheo} has a counterpart in the super--p--brane formulation considered
below, but we have tried to expose the results in a selfcontained form, so
that a special knowledge of the group--manifold approach is not
required.

The idea of applying the generalized action principle to considering
super--p--branes emerged due to the following reason.

In recent years much attention has been payed to finding the origin and
the geometrical meaning of the local fermionic $\k$--symmetry \cite{k} of
super--p--branes in the Green--Schwarz formulation \cite{gs}
with the aim to advance in solving the problem of the covariant
quantization of superstrings. This resulted in the development of
different versions of a
twistor--like approach \cite{stv}--\cite{ghs}.
In a lorentz-harmonic twistor--like formulation of refs.
\cite{bh}--\cite{bzm1}
the $\k$--symmetry was represented in an irreducible but ruther
complecated form. The twistor--like approach based on a superfield
formulation of super--p--branes in world superspace
\cite{stv}--\cite{bers94} allowed one
to replace the $\k$--symmetry by more fundamental local world
supersymmetry and thereby to solve the problem of the infinite
reducibility of the former. This revealed a variety of new interesting
features in describing the dynamics and elucidating the world
geometry of the super--p--branes \cite{stv}--\cite{bpstv}.
In particular, the natural
appearance of twistor variables, or Lorentz harmonics, gave rise to
geometrical problems of embedding a supersurface into a target
superspace, which predetermined the structure of the super--p--brane
action.

At the same time some basic problems
have not been solved satisfactory in the known versions of the
approach both from the
aesthetic and practical point of view. For instance, for constructing
the superfield action one should use superfield Lagrange multipliers.
Though some
of their components can be identified (on the mass shell) with the
momentum density and the tension of the super--p--brane, in general, the
geometrical and physical meaning of the Lagrange multipliers is obscure.
Moreover, in a version suitable for the description of D=10, 11 objects
\cite{to,dghs92,gs2,tp93,bers94}
 their presence in the action gives rise to some new symmetries
which turn out to be infinite reducible themselves, so that the problem
which we fighted in the conventional Green--Schwarz formulation
reappeared in a new form in the twistor--like formulation. Another point
concerning the Lagrange multipliers is that in the superfield
formulation of D=10 type II superstrings \cite{gs2} and a D=11, N=1
supermembrane \cite{tp93} Lagrange multipliers become propagative redundant
degrees of freedom which may spoil the theory at the quantum level.

All this forces one to revise the twistor--like superfield approach and
to find its more geometrically grounded version. To this end we have
turned to the generalized action principle of the rheonomic
approach.

Our notation and convention are as follows.
The small Latin indices stand
for vectors and the Greek indices stand for spinors. All underlined
indices correspond to target superspace of D bosonic dimensions, and
that which are not underlined correspond to super--p--brane
world supersurface.
The indices from the beginning of the
alphabets denote the vector and spinor components in the tangent
superspace. Indices from the second half of the alphabets are world
indices:
$$
{\underline a},{\underline b},{\underline c}=0,...,D-1\qquad
{\underline l},{\underline m},{\underline n}=0,...,D-1;
$$
$$
a,b,c=0,...,p \qquad l,m,n=0,...,p
$$
$$
{\underline \a},{\underline \b},{\underline \g}=1,...,2^{[{D\over 2}]}
{}~~(or~2^{[{D\over 2}-1]})\qquad
{\underline \mu},{\underline \nu},{\underline \rho}=1,...,2^{[{D\over 2}]}
{}~~(or~2^{[{D\over 2}-1]});
$$
$$
\a,\b,\g=1,...,2^{[{{p+1}\over 2}]}\qquad
\mu,\nu,\rho=1,...,2^{[{{p+1}\over 2}]}
$$
$
i,j,k=1,...,D-p-1$~~{\rm{stand~for~ the ~vector ~representation~ of}~
$SO(D-p-1)$};\\
$
p,q,r~(or~\dot p,\dot q,\dot r)=1,...,D-p-1$~~~{\rm{stand~for~ a spinor
 ~representation~ of}~ $SO(D-p-1)$}.

Target superspace is parametrized by bosonic coordinates $X^{\underline{m}}$
and fermionic coordinates $\Th^{\underline{\mu}}$, and world superspace
is parametrized by bosonic coordinates $\xi^m$ and fermionic coordinates
$\eta^{\a p}$.
The number of  $\eta^{\a p}$ is to be half the number of
 $\Th^{\underline{\mu}}$.
This ensures that all independent
$\k$--symmetry transformations are replace by the world supersurface
diffeomorphisms.

\section{The generalized action principle for super--p--branes}

The super--p--brane formulation considered below is based on the
following basic principles akin to the rheonomic approach of refs.
\cite{rheo}, however our case is much more simple since for
constructing the action only the simplest geometrical objects, i.e.
vielbeins, and not connection and curvature are involved:
\begin{description}
\item[i)]
The action is a differential (p+1)--superform integrated over a
(p+1)--dimensional bosonic submanifold ${\cal
M}_{p+1}~:\{(\xi^m,\eta);~\eta=\eta(\xi)\}$ on the world supersurface
$$
S=\int\limits^{}_{{\cal M}_{p+1}}L_{p+1}.
$$
The Lagrangian $L_{p+1}$ is constructed  out of vielbein
differential one--forms
in target superspace and world supersurface ({\sl a priori} considered
as independent) by use of exterior product
of the forms without any application of the Hodge operation, for this
only even world supersurface vielbeins are used, thus $\xi$--directions
have a privilege over $\eta$--directions.

To get the superfield equations of motion both the coefficients of
the forms and the bosonic submanifold are varied.
The variation of the action over ${\cal M}_{p+1}$ is
amount to superdiffeomorphism transformations on the world supersurface.
%$\left(\xi,\eta(\xi)\right)~\rightarrow~\left(\xi^\prime(\xi,\eta(\xi)),
%\eta^\prime(\xi,\eta(\xi))\right)$.
%%%%%%
%and is reduced to the following
%integral (provided ${\cal M}_{p+1}$ is closed)
%$$
%\d_{{\cal M}_{p+1}}S=
%\int\limits^{}_{{\cal M}_{p+1}}dL_{p+1}(d_1,...,d_{p+1},\d),
%$$
%where in the (p+2)--form $dL_{p+1}$ one of the external differentials is
%replaced by $\d=\d\xi^m\partial_m+\d\eta^{\a p}\partial_{\a p}$.
This allows one to extend the superfield equations from ${\cal M}_{p+1}$
to the whole world supersurface.
\item[ii)]
The intrinsic geometry of the world supersurface is not {\sl a priori}
restricted by any superfield constraints, and the embedding of the world
supersurface into target superspace is not {\sl a priori} specified by
any condition such as a geometrodynamical condition \cite{stv}--\cite{bpstv}
(see eq.
\p{gd}) the latter playing the crucial role in the twistor--like
superfield approach. All the constraints and the geometrodynamical
condition are obtained as equations from the action constructed with the
generalized action principle.
\item[iii)]
The field variation of the action gives two kinds of relations:\\
{}~~~1) relations between target superspace and world
supersurface vielbeins which orientate them along one another and
are the standard relations of surface
embedding theory; we call them ``rheotropic'' conditions
\footnote{`rheo' is `current' and `tropos' is `direction, rotation' in
Greek}; \\
{}~~~2) dynamical equations causing the embedding to be minimal.\\
Only the latter equations put the theory on the mass shell.
%Thus in the same way as in the group--manifold approach only the bosonic
%directions along the world supersurface are horizontal from the fiber
%bundle point of view; the bosonic directions orthogonal to the world
%supersurface and the fermionic directions along the world supersurface
%are treated as vertical ones and are completely determined through the
%horizontal directions (note that the physical degrees of freedom of the
%super--p--brane are
%described just by $X^{\underline{m}}$ and $\Th^{\underline{\mu}}$
%orthogonal to the world supersurface).
\item[iv)]
The theory is superdiffeomorphism invarinat off the mass shell if for
the action \p{1} to be independent of the surface ${\cal M}_{p+1}$ (i.e.
$dL_{p+1}=0$) only the rheotropic relations are required, and the latter
do not lead to equations of motion.

Upon eliminating auxiliary fields
one reduces the superdiffeomorphism
transformations of the superfields of the model to that of the $\k$--symmetry.
\end{description}

With all these points in mind we propose a super--p--brane action in the
following form:
\begin{eqnarray}\label{1}
S_{ D,p}& =-{{(-1)^p}\over{p!}}
\int\limits^{}_{{\cal M}_{p+1}}
\left(E^{a_0}
e^{a_1}...e^{a_{p}}\varepsilon_{a_0a_1...a_{p}}
-{{p}\over{(p+1)}}
e^{a_0} e^{a_1}... e^{a_{p}}\varepsilon_{a_0a_1...a_{p}}\right) \nn
&\pm(-i)^{{{p(p-1)}\over 2}-1}{p\over{(p+1)!}}
\int\limits^{}_{{\cal M}_{p+1}}
\sum^{p+1}_{k=0}
\Pi^{\underline{m}_p} \ldots
\Pi^{\underline{m}_{k+1}} dX^{\underline{m}_k}
\ldots dX^{\underline{m}_1}
d\Theta \Gamma_{\underline{m}_1\ldots \underline{m}_p}
\Theta
\end{eqnarray}
where the wedge product of the differential forms is implied,
$\varepsilon_{a_0a_1...a_{p}}$ is the unit antisymmetric tensor on
${\cal M}_{p+1}$, and p--brane tension is chosen to be one.

In \p{1}
$e^a(\xi,\eta)$ are the bosonic vector components of a world supersurface
vielbein one--form $e^A=(e^a,e^{\a p})$,
then the external differential $d$ can be
expended in the $e^A$ basis as follows
\begin{equation}\label{d}
d=e^aD_a+e^{\a p}D_{\a p}
\end{equation}
with $D_a,~D_{\a p}$ being world--supersurface covariant derivatives.
\begin{equation}\label{pif}
\Pi^{\underline{m}}=dX^{\underline{m}}-id\Th\G^{\underline{m}}\Th,\qquad
d\Th^{\underline{\mu}}
\end{equation}
is the pullback onto world supersurface of the basic supercovariant forms
in flat target superspace.
$u_{\underline{m}}^{a}(\xi,\eta)$ are (p+1) vector components
of a local frame (supervielbein)
\begin{equation}\label{E}
E^{\underline a}=
\Pi^{\underline m}u_{\underline m}^{\underline a},~E^{\underline \a}
= d\Th^{\underline \mu}v_{\underline \mu}^{\underline \a}
\end{equation}
in target superspace. Together with the (D--p--1) components
$u_{\underline{m}}^{i}$ they are naturally \cite{bh}--\cite{bzm1,ghs,bpstv}
composed of
the spinor
components (Lorentz harmonics) $v_{\underline{\mu}}^{\underline{\a}} =
( v_{\underline{\mu}\a q}~,~v^{\a}_{\underline{\mu}\dot q} )$
 of the local frame:
\begin{equation}\label{2}
\d_{qp} (\gamma_a )_{\a \b}
u^{~a}_{\underline{ m}} =
v_{\a q}
\Gamma_{\underline{ m}}
v_{\b p} ,
\qquad
\d_{\dot{q} \dot{p}} (\gamma_{a})^{\a \b}
u^{a}_{\underline{ m}} =
v^{\a }_{\dot q}
\Gamma_{\underline{m}}
v^{\b }_{\dot p} ,
\qquad
\d^{\a}_{\b} \gamma ^{i}_{q \dot p}
u^{i}_{\underline{ m}}
= v_{\a q}
\Gamma_{\underline{ m}}
v^{\b }_{\dot p}.
\end{equation}
where $\G_{\underline{m}}$, $\gamma_{a}^{\a \b}$ and
$\gamma ^{i}_{q \dot p}$
 are the $SO(1,D-1)$, $SO(1,p)$ and
 $SO(D-p-1)$  $\g$--matrices, respectively,
\begin{equation}\label{gam}
(\Gamma ^{a})_{\underline{\b}\underline\a}
={\it diag} \left( \gamma ^{a}_{ \b\a} \delta _{qp},
- \gamma ^{a}_{\a\b} \delta _{\dot{q} \dot{p}} \right),
\qquad
(\Gamma ^{i})_{\underline{\b}\underline\a}=
\left( \matrix{ 0  ~~\epsilon _{\a \b} \gamma ^{i}_{{q}\dot{p}} \cr
- \epsilon _{\a \b} \tilde{\gamma}^{i}_{{\dot q}{p}}~~0 \cr}
\right).
\end{equation}

  The matrix $v_{\underline{\mu}}^{~\underline{\a}}$ takes it values in
$Spin(1,D-1)$, and its components (as well as
$u_{\underline{m}}^{\underline{a}}=(u_{\underline{m}}^{{a}},
u_{\underline{m}}^{i}$)) parametrize a coset space
${SO(1,D-1) \over {SO(1,p)\times SO(D-p-1)}}$. Note that
\begin{equation}\label{o}
u^{~\underline{a}}_{\underline{m}}\eta^{\underline{m}\underline{n}}
u_{\underline{n}\underline{b}} = \eta^{\underline{a}\underline{b}}=
{\it diag}(1,-1,\ldots ,-1)
\end{equation}
(see \cite{gikos,sok,bh}--\cite{bzm1,ghs,bpstv}
for the details on the  harmonics).

As we will see below, the rheotropic conditions cause the target superspace
vielbein \p{E} components $E^a,~E^{\a p}$ to become
tangent and $E^i$ to become
orthogonal to the world supersurface.

When the superfields are restricted to their leading components
(i.e. at $\eta=0$) and  in \p{d} only
the vector components are taken into account,
eq. \p{1} is reduced to a component
super--p--brane action considered earlier in \cite{bh}--\cite{bzm1}
the latter being
classically equivalent to the conventional Green--Schwarz formulation.
The new fundamental
feature of \p{1} is that it is constructed solely out of the differential
forms \cite{bpstv}. \footnote{The same situation one encountered in
the case of N=1
supergravity in the group--manifold approach, where upon matching all
the constant parameters in a supergravity action to satisfy rheonomic
requirements \cite{rheo} one recovers the action written in terms of
differential forms which was firstly discovered in \cite{vs} and
rediscovered in \cite{dz}.}
The last term in \p{1} is the Wess-Zumino term \cite{gs},
its coefficient being
fixed by the requirement that when the action \p{1} is restricted to the
component formulation of the super--p--branes \cite{gs,bzst1,bzm1}
the resulting
action has local $\k$--symmetry \footnote{from the rheonomy point of
view the value of the coefficient is fixed by the requirement that field
equations obtained from the action reproduce the rheotropic conditions}.
As to the superfield action \p{1}
itself, it  {\sl does not
possess} $\k$--symmetry in its conventional form.

\section{Equations of motion}

Varying \p{1} over $u^{a}_{\underline{m}}$ (with taking into
account \p{o}), over $e^a$, $X^{\underline{m}}$ and $\Th^{\underline{\mu}}$ we
get the following differential form equations
\begin{equation}\label{u}
{\d S\over{\d u^{a}_{\underline{m}}}}\Rightarrow\Pi^{\underline{m}}
u_{\underline{m}}^{i}
e^{a_1}...e^{a_{p}}\varepsilon_{a_0a_1...a_{p}}=0,
\end{equation}
\begin{equation}\label{e}
{\d S\over{\d e^a}}~\Rightarrow~(\Pi^{\underline{m}}
u_{\underline{m}}^{a_0}
-e^{a_0})e^{a_1}...e^{a_{p-1}}\varepsilon_{a_0a_1...a_{p-1}a}=0,
\end{equation}
\begin{equation}\label{x}
{\d S\over{\d X^{\underline{m}}}}~\Rightarrow~d(u_{\underline{m}}^{a_0}
e^{a_1}...e^{a_{p}}\varepsilon_{a_0a_1...a_{p}})
\pm (-1)^p(-i)^{{p(p-1)}\over
2}\Pi^{\underline{m_p}}...\Pi^{\underline{m_2}}d\Th
\G_{\underline{m}\underline{m}_2...\underline{m}_p}d\Th=0,
\end{equation}
\begin{equation}\label{th}
{\d S\over{\d\Th^{\underline{\mu}}}}~\Rightarrow~
d\Th^{\underline{\mu}}\G^{\underline{m}}_{\underline{\mu\nu}}
u_{\underline{m}}^{a_0}
e^{a_1}...e^{a_{p}}\varepsilon_{a_0a_1...a_{p}}
\pm (-1)^p(-i)^{{p(p-1)}\over
2}d\Th^{\underline{\mu}}
(\G_{\underline{m}_1...\underline{m}_p})_{\underline{\mu\nu}}
\Pi^{\underline{m_p}}...\Pi^{\underline{m_1}}=0.
\end{equation}

{}From \p{u} and \p{e} we get part of the rheotropic conditions
\begin{equation}\label{pi}
\Pi^{\underline{m}}u_{\underline{m}}^a=e^a,\qquad
\Pi^{\underline{m}}u_{\underline{m}}^i=0 \qquad \Rightarrow \qquad
\Pi^{\underline{m}}=e^au^{\underline{m}}_a,
\end{equation}
while from \p{th}, \p{2} and \p{pi} it follows that \cite{bzm1}
\begin{equation}\label{th1}
\varepsilon_{aa_1...a_{p}}e^{a_1}...e^{a_{p}}
d\Th^{\underline{\mu}}v_{\underline{\mu}\a\dot q}(\g^a)^\a_\b=0,
\end{equation}
which results in
\begin{equation}\label{th1c}
(\g^a)^\a_\b D_a \Th^{\underline{\mu}}
v_{\underline{\mu}\a\dot q} = 0 ,
\end{equation}
\begin{equation}\label{th1s}
D_{\a p} \Th^{\underline{\mu}} v_{\underline{\mu}\b\dot q} = 0.
\end{equation}
Eq. \p{th1c} is a dynamical equation of motion, while \p{th1s}
belongs to the rheotropic conditions.

One can directly check that for the Lagrangian in \p{1} to be a closed
differential form it is sufficient that only the rheotropic conditions
\p{pi}, \p{th1s} are valid,
which in its own turn ensures the equations of motion \p{u}--\p{th}
to be valid on the whole world supersurface (see item {\bf i)}
of the generalized action
principle). As is well known \cite{stv}--\cite{bpstv}, for the case
of N=1 superparticles and N=1 heterotic strings
in D=3,4,6 and 10, as well as N=2 superstrings in D=3 eqs. \p{pi},
\p{th1s} do not lead to the dynamical equations of motion  and  allow
for the models to be superdiffeomorphism invariant off the mass shell
(see items {\rm iii), iv)} of the generalized action principle).

However, in the case of N=1, D=11 supermembrane and N=2, D=10
superstrings eq. \p{th1s} results in the equation of motion \p{th1c}, which
holds the theories on the mass shell.

\section{Component formulation}

The component formulation \cite{bh}--\cite{bzm1,bpstv}
of super--p--branes is obtained by
choosing the surface ${\cal M}_{p+1}$ to be defined by the condition
$\eta^{\a q}=0$ and taking  into account only the vector components of
\p{d}.

For $X^{\underline m}|_{\eta=0}=x^{\underline m}(\xi)$,
$\Th^{\underline \mu}|_{\eta=0}=\th^{\underline \mu}(\xi)$,
$u^{\underline{m}}_a|_{\eta=0}=u^{\underline{m}}_a(\xi)$ and
$e^a|_{\eta=0}=e^a(\xi)$ one can get from \p{u}--\p{th1}, \p{th1c}
the following equations:
\begin{equation}\label{pic}
\Pi_a^{\underline{m}}=e_a^m(\partial_m x^{\underline{m}}
-i\partial_m\th\G^{\underline{m}}\th)=
u^{\underline{m}}_a(\xi),
\end{equation}
\begin{equation}\label{thc}
(\g^a)^\a_\b e_a^m\partial_m\th^{\underline{\mu}}
v_{\underline{\mu}\a\dot q}(\xi)=0,
\end{equation}
\begin{equation}\label{xc}
\partial_m(\sqrt{-g}g^{mn}\Pi^{\underline m}_n)
\pm (-i)^{{p(p-1)}\over 2}
\varepsilon_{m_p...m_{0}}
\Pi^{\underline{m}_p}_{m_p}...\Pi^{\underline{m}_2}_{m_2}
\partial_{m_1}\th
\G_{\underline{m}\underline{m}_2...\underline{m}_p}\partial_{m_0}\th=0,
\end{equation}
where $g_{mn}=e_m^ae_{an}=\Pi^{\underline m}_m\Pi_{{\underline m}n}$
is the induced metric on the world surface.

Eq. \p{xc} is the same as one obtains in the standard Green--Schwarz
formulation \cite{gs} where the variations over $g_{mn}$ are equivalent
to that of $e^a_m$ herein.

Note once again that the component action obtained by restricting \p{1}
to the leading components of the superfields, as well as eqs.
\p{pic}--\p{xc}, does possess the $\k$-symmetry
in an irreducible form \cite{bzst1,bzm1}.
%\begin{equation}\label{k}
%\d\th^{\underline\mu}=\k^{\a p}(\xi)v^{\underline\mu}_{\a p},
%\qquad
%\d x^{\underline m}=-i\th\G^{\underline m}\d\th,
%\qquad
%\d e^a =2i \k^{\a p}\g^a_{\a\b}v_{\underline{\mu}\b
%p}d\th^{\underline\mu},
%\end{equation}
%\begin{equation}\label{dv}
%\d v^{\underline\mu}_{\a p} \sim
%det(e_a^m)i\k^\b_p\g^a_{\b\a}e^m_a\partial_m\th^{\underline\mu}
%+~(terms~proportional~ to~ eqs.~of~motion~of~ \th)
%\end{equation}
The basic feature of the twistor--like superfield approach is that the
$\k$--symmetry transformations  are the relic of the world surface
superdiffeomorphisms \cite{stv}--\cite{bers94},
for instance, $\th^{\underline\mu}$ and
$v^{\underline\mu}_{\a p}$ are transformed as superpartners
(where $v^{\underline{\mu}}_{\underline\a}=
(v^{\underline{\mu}}_{\a\dot q},v^{\underline{\mu}}_{q\a})$ is
inverse of $v_{\underline{\mu}}^{\underline\a}$.)
%\footnote{It is just because the transformations \p{dv}
%of $v^{\underline\mu}_{\a p}$
%cannot be written in terms of differential forms it is not possible to
%extend the $\k$--symmetry
%transformations \p{k}--\p{dv} to that of the superfields in such a
%way that the superfield action \p{1} has this symmetry as an independent
%one.}

Thus we conclude that the formulation under consideration reproduces the
conventional versions of the super--p--branes.

\section{Constraints on the world supersurface geometry induced by
embedding}

Let us analyze the consequences of the superfield equations
\p{u}--\p{th1}.

{}From \p{e} we obtain the rheotropic condition
\begin{equation}\label{ea}
e^a=\Pi^{\underline m}u_{\underline m}^a=E^a,
\end{equation}
which means that on the mass shell the form of $e^a$ is induced by
embedding and is determined by the pullback of target superspace vector
vielbein components.

Then we get
\begin{equation}\label{gd}
\Pi^{\underline m}_{\a p}=D_{\a p}X^{\underline m}-
iD_{\a p}\Th\G^{\underline m}\Th=0,
\end{equation}
\begin{equation}\label{tw}
\Pi^{\underline m}_a=D_aX^{\underline m}-iD_a\Th\G^{\underline m}\Th=
u^{\underline m}_a.
\end{equation}
Eq. \p{gd} is the geometrodynamical condition on the embedding of the
world supersurface, and \p{tw} is the twistor--like solution to the
Virasoro--like constraints $\Pi^{\underline
m}_a\Pi_{\underline{m}a}=\eta_{ab}$
on the dynamics of the super--p--branes \cite{stv}--\cite{bzm1}.
The latter is connected
with  the former through the consistency requirements the general
expression for them being:
\begin{equation}\label{spi}
d\Pi^{\underline m}=-id\Th\G^{\underline m}d\Th=T^au^{\underline m}_a
+e^aDu^{\underline m}_a,
\end{equation}
where $T^a=De^a\equiv de^a-\Om^a_be^b$ is the world surface torsion and
$$
\Om^{ab}=u^{a\underline m}du_{\underline m}^b~~~\Rightarrow~~~
u^{a\underline m}Du_{\underline m}^b\equiv 0,
$$
\begin{equation}\label{con}
\Om^{ij}=u^{i\underline m}du_{\underline m}^j~~~\Rightarrow~~~
u^{i\underline m}Du_{\underline m}^j\equiv 0.
\end{equation}
 is the $SO(1,p)\times SO(p-1)$
connection induced by the embedding.
{}From \p{spi}, \p{2} we see that $T^a$ is constrained by the embedding to be
\begin{equation}\label{tor}
T^a=-id\Th\G^{\underline m}d\Th u_{\underline m}^a=
-id\Th^{\underline\mu}v_{\underline\mu}^{\underline\a}\G^{a}_{\underline{\a\b}}
d\Th^{\underline\nu}v_{\underline\nu}^{\underline\b}.
\end{equation}
Let us turn to eq. \p{th1s}.
It has the general solution
\begin{equation}\label{Th3}
D_{\a p}\Th^{\underline{\mu}}=A_{\a p}^{\b q}
v^{\underline{\mu}}_{\b q},
\end{equation}
where $A_{\a p}^{\b q}(\xi,\eta)$ is a matrix.

Now note that since the spinor components $e^{\a p}$ of the world
supersurface vielbein and the intrinsic connection form
are not involved into the
construction of the action \p{1}, the form of the action
admits the following redefinition of $e^{\a p}$
and the corresponding redefinition of $D_{\a p}, D_a$:
$$
e^{\a p}~~\rightarrow~~(e^{\b p} + e^b\chi_b^{\b q})A_{\b q}^{\a p},
$$
\begin{equation}\label{trans}
D_{\a p}~~\rightarrow~~(A^{-1})_{\a p}^{\b q}D_{\b q},~~
D_a~~\rightarrow~~D_a-\chi_a^{\b q}D_{\a p},
\end{equation}
where $A_{\b q}^{\a p}(\xi,\eta)$
is a nonsingular matrix, and $\chi_a^{\b q}(\xi,\eta)$ are Grassmann
``boosts''.

By use of the A--transformations in \p{trans} we can reduce \p{Th3}
to
\begin{equation}\label{Th4}
D_{\a p}\Th^{\underline{\mu}}=v^{\underline{\mu}}_{\a p}.
\end{equation}
Eq.\p{Th4} identifies $D_{\a p}\Th^{\underline{\mu}}$ with Lorentz
harmonics (twistor--like variables) \cite{ghs,bpstv}.

The remaining $\chi$--transformations in \p{trans} can be used to put
\begin{equation}\label{Th5}
D_{a}\Th^{\underline{\mu}}v_{\underline{\mu}\a p}=0.
\end{equation}

{}From \p{Th4}, \p{Th5} it follows that, as $e^a$ (eq. \p{ea}), the spinor
components $e^{\a p}$ of the world supervielbein are related to
the pullback of target superspace spinor
vielbein components by the rheotropic condition
\begin{equation}\label{eal}
e^{\a p}=d\Th^{\underline{\mu}}v_{\underline{\mu}}^{\a p}=E^{\a p}.
\end{equation}

Substituting eq. \p{eal} into \p{tor} we obtain the most essential
torsion constraint
\begin{equation}\label{tv}
T^a_{\a p\b q}=-2i\g^a_{\a\b}\d_{pq}
\end{equation}
usually imposed in the superfield formulations of supergravity. In
particular, in the case of super--p--branes \cite{stv}--\cite{bpstv}
all other torsion
constraints
%,which are obtained by solving the Bianchi identities,
are conventional ones, and a definite set of the torsion
constraints can be chosen
by redefining the vielbein and connection forms
\cite{con} (as in \p{trans}).
For instance, for our choice of $e^{\a p}$ (eq. \p{eal}) it
also follows that:
\begin{equation}\label{ts}
T^{\a p}_{\b q\g r}=0.
\end{equation}

\section{Conclusion}

Applying the generalized action principle of the rheonomic approach
\cite{rheo}
to describing super--p--branes allowed us to construct the superfield
action \p{1} of the differential vielbein superforms on the world
supersurface and in target superspace without any use of Lagrange
multipliers. This allowed us to escape the problem of infinite
irreducible symmetries and redundant propagating degrees of freedom.

In contrast to the conventional superfield approach none restrictions
have been imposed by hand on the geometry of the world
supersurface and on its embedding into the target superspace. The
geometrodynamical condition \p{gd} on the embedding, the twistor--like
constraint \p{tw}, the induced form of the world supersurface vielbeins
\p{ea}, \p{eal} and connections \p{con}, as well as the torsion
constraints on world--surface supergravity \p{tv}, \p{ts} have arisen as
consequences of differential form equations (rheotropic conditions) obtained
from the action \p{1}.

The action also provides  the superfield equations of motion \p{x},
\p{th} (or \p{th1c}) of the super--p--brane in a form suitable for
developing the geometrical approach \cite{geo} to
describe super--p--branes \cite{bpstv}.

When restricted to the leading components of the superfields
$X^{\underline m}$, $\Th^{\underline\mu}$,
$v^{\underline\mu}_{\underline\a}$ and $e^a$ the equations of motion
coincide with that of the conventional formulations  and are
invariant under the irreducible $\k$--symmetry transformations
being the relic of the world supersurface diffeomorphisms.

We stress that the prescription proposed in the present article is
valid for the doubly supersymmetric formulation of super--p--branes in
space--time of any number of dimensions suitable for their existence
\cite{gs}, and can be generalized to a curved background.

In the case of D=3,4,6,10 heterotic strings, and N=2, D=3 superstrings,
where the rheotropic conditions \p{pi}, \p{th1s} do not lead to dynamical
equations of motion one may try to think of how to covariantly quantize the
theory on the ground of the approach proposed herein.

\bigskip
{\bf Acknowledgements}. D.S. is grateful to Paolo Pasti and Mario Tonin
for illuminating questions and fruitful discussion.

\end{document}